# Weak ferromagnetism in non-centrosymmetric BiPd 4K superconductor

Rajveer Jha[1], Reena Goyal[1], P. Neha[2], V. Maurya[2], A. Srivastava[1], Anurag Gupta[1], S. Patnaik[2] and V.P.S. Awana[1*]

[1]CSIR-National Physical Laboratory, Dr. K. S. Krishnan Marg, New Delhi-110012, India
[2]School of Physical Sciences, Jawaharlal Nehru University, New Delhi-110067, India

**ABSTRACT**

We report synthesis of non-centrosymmetric BiPd single crystal by self flux method. The BiPd single crystal is crystallized in monoclinic structure with the $P2_1$ space group. Detailed SEM (Scanning Electron Microscopy) results show that the crystals are formed in slab like morphology with homogenous distribution of Bi and Pd. The magnetic susceptibility measurement confirmed that the BiPd compound is superconducting below 4K. Further, BiPd exhibits weak ferromagnetism near the superconducting transition temperature in isothermal magnetization (MH) measurements. The temperature dependent electrical resistivity also confirmed that the BiPd single crystal is superconducting at $T_c=4K$. Magneto transport measurements showed that the estimated $H_{c2}(0)$ value is around 7.0kOe. We also obtained a sharp peak in heat capacity $C_p(T)$ measurements at below 4K due to superconducting ordering. The normalized specific-heat jump, $\Delta C/\gamma T_c$, is 1.52, suggesting the BiPd to be an intermediate BCS coupled superconductor. The pressure dependent electrical resistivity shows the $T_c$ decreases with increasing applied pressure and the obtained $dT_c/dP$ is -0.62K/Gpa.

Key words: Noncentrosymmetric superconductors, BiPd crystal growth, Electrical, Magnetic characterization

**\*Corresponding Author**
Dr. V. P. S. Awana,
Principal Scientist
E-mail: awana@mail.npindia.org
Ph. +91-11-45609357, Fax-+91-11-45609310
Homepage awanavps.wenbs.com

**INTRODUCTION**

The discovery of superconductivity in noncentrosymmetric heavy fermion superconductors viz. $CePt_3Si$ [1] has attracted great attention of condensed matter physics community. Similarly, some other noncentrosymmetric superconductors have been studied in recent past with conventional superconducting behavior such as $Ru_7B_3$, $Mg_{10}Ir_{19}B_{16}$, α-BiPd, and the cubic phases of $Re_3W$, and $Re_{24}Nb_5$ [2-7]. Further, superconductivity has been reported in n-type metallic interface being formed between the $LaAlO_3$ and $SrTiO_3$ interface, which is naturally an inversion symmetry system without up-down reflection [8]. The name "noncentrosymmetric" distinguish the symmetry of a crystal lattice lacking an inversion center. In such materials, the parity allows explicitly the discerning between spin-singlet and spin-triplet pairs in the superconducting phase [9]. Moreover, Anderson suggested the center of inversion in the crystal structure as an essential symmetry element for spin-triplet pairing [10]. On the other hand, the time-reversal symmetry is required for the spin-singlet Cooper pairing [11]. It has also been proposed that the noncentrosymmetric crystal structure along with strong correlation between electrons may lead to the mixing of spin singlet and spin-triplet superconductivity [9]. In addition, various physical properties for example thermal conductivity and magnetic penetration depth proposed line node in the energy gap [12-16]. In some theoretical studies the exotic mixed-state phases have been predicted for this type of superconductors. In fact theoretical studies predict heavy fermion as well as the unconventional Cooper pairing for these new superconducting materials [17]. Interestingly, yet the experimental studies on noncentrosymmetric materials did not displayed heavy fermion features and rather conventional superconductivity with small spin-orbit scattering is seen [2-7]. A recent scanning tunnelling microscopy (STM) study on noncentrosymmetric BiPd superconductor confirmed the same to be a single gap, s-wave superconductor [18].

Keeping in view, the importance of noncentrosymmetric superconducting materials, in this work, we report the synthesis and superconductivity characterization of noncentrosymmetric $\alpha$-BiPd single crystalline material. The single crystals of $\alpha$-BiPd were obtained by self flux melt growth method, which are well crystallized in monoclinic structure with space group $P2_1$. The temperature dependence of magnetic susceptibility and electrical resistivity confirmed the $\alpha$-BiPd compound to be superconducting ($T_c$) at below 4K, which is slightly higher than the earlier reported value of 3.8K [5]. Further, we observed week ferromagnetism at just close to the $T_c$ i.e., at 4K in isothermal magnetization measurements. Heat capacity measurements for the studied single crystal BiPd suggested the same to be an intermediate BCS coupled superconductor. The electrical resistivity under hydrostatic pressure studies for single crystal BiPd suggested negative pressure coefficient of $dT_c/dP = -0.62$K/Gpa.

**EXPERIMENTAL DETAILS**

The investigated crystals of $\alpha$-BiPd were grown by a self flux melt growth method. The crystals have a platelet like shape with shiny surfaces. Initially, we take high purity (99.99%) Pd and Bi powders, weighed them in stoichiometric ratios and grind thoroughly in the argon filled glove box. The mixed powder is afterward pelletized by applying uniaxial stress of 100 kg/cm$^2$ and then pellet was sealed in an evacuated ($<10^{-4}$ Torr) quartz tube. The quartz ampoule was put in the furnace at temperature of 650$^o$C with a rate of 2$^o$C/min for 12h and then the furnace is cooled slowly at a rate of 1$^o$C/minute down to room temperature. We performed room temperature X-ray diffraction (XRD) for the obtained BiPd crystal on Rigaku x-ray diffractometer using CuK$_\alpha$ line of 1.54Å. The morphology of the obtained single crystal has been seen by scanning electron microscopy (SEM) images on a ZEISS-EVO MA-10 scanning electron microscope, and the Energy Dispersive X-ray spectroscopy (EDAX) was employed for elemental analysis. The magnetic measurements were carried out on SQUID magnetometer MPMS 7Tesla, Quantum Design, USA. The temperature-dependence of the magnetic susceptibility $\chi$ is measured in an applied field of 10 Oe for temperature range from 1.8 to 7K. Transport measurements were carried out on Physical Property Measurements System (PPMS-14Tesla-Quantum Design). The pressure dependent electrical resistivity was measured using the HPC-33 Piston type pressure cell with Quantum design DC resistivity option on existing PPMS-14Tesla.

**RESULTS AND DISCUSSION**

Figure 1 shows the XRD pattern of crushed powder of single crystalline BiPd compound. BiPd compound is crystallized in monoclinic structure with the $P2_1$ space group. The estimated values of the lattice constants are $a = 5.62(1)$Å, $b = 10.66(2)$ Å, $c = 5.67(1)$ Å, with $\alpha = \gamma = 90°$ and $\beta = 101^0$. The structure consists of 16 atoms per unit cell having four equivalent sites for Bi and another four equivalent sites for Pd. The unit cell contains alternate layer of Bi and Pd sheets with short Pd-Pd distances without inversion symmetry in the structure. The schematic unit cell of BiPd is shown in inset of Fig. (1). Figure 2 shows the SEM image of BiPd single crystalline compound; here one can clearly see that the single crystal is grown in the layered form. The EDAX analysis, being shown adjacent to the SEM image in Fig. 2, established that the Bi and Pd chemical elements in BiPd crystal are in the stoichiometric ratio.

Detailed micro-structural characterization of BiPd alloy was carried out using high resolution transmission electron microscope (HRTEM, model: Tecnai G2 F30 STWIN assisted with the field emission gun for the electron source at an electron accelerating voltage of 300 kV). Several interesting features in the real and reciprocal space are delineated on this alloy. In general a uniform microstructure was observed throughout in the bright field electron micrographs recorded at different magnifications (Fig. 3 a,b). The grey level contrast in the microstructure is evident for a highly crystalline material with the grain boundaries at instances. One such grain boundary has been marked with a set of arrows between the grains 1 and 2 in Fig. 3b. At atomic scale it was revealed that the grain boundary is possibly contrast evolved presumably due to minor compositional variation in the microstructure, otherwise the crystallographic planes with an inter-planar spacing of 0.27nm and hkl: $\bar{2}\,12$ (crystal structure: monoclinic, space group: $P2_1$, lattice parameters: a=0.72nm, b=1.07nm, c=0.87nm, β=89.7°, reference: JCPDS card no. 330213) are mostly continuous in this perturbed region too (inset in Fig. 3b). A few ultra-fine inclusions of Pd are also elucidated in the matrix of BiPd, as marked with a set of double arrows in Fig. 3b. These Pd inclusions are presumably precipitated during synthesis of the alloy and are distributed occasionally in a random manner in the matrix of BiPd. At lattice scale it is clearly depicted that the atomic planes with the inter-planar spacing of 0.34 nm and hkl: 022 of monoclinic BiPd are stacked throughout in the matrix (Fig. 3c), whereas, as an illustrative example, an inclusion Pd (size ~ 8 nm) is constituted of planes with inter-planar spacing of 0.23 nm with hkl: 111 (crystal structure: fcc, space group: $Fm\bar{3}m$, lattice parameter: a=0.39 nm, reference: JCPDS card no. 894897). A single crystal electron diffraction pattern, recorded from BiPd matrix, exhibited spotty pattern along $[1\bar{1}0]$ zone i.e., axis of a monoclinic crystal structure in reciprocal space. Two important planes with hkl: 004 and $2\bar{2}0$, are marked in Fig. 3d Another electron diffraction pattern from a composite of the BiPd matrix encompassing fine Pd inclusions show the single crystal pattern of a monoclic crystal of BiPd along $[1\bar{2}\bar{3}]$ zone axis, and faint Debye rings evolved from different atomic planes of a fcc Pd crystal. Some of the important atomic planes of fcc Pd corresponding to inter-planar spacings of 0.23, 0.14, 0.12 nm with hkl: 111, 220, 311, are marked as a,b,c on Debye rings, respectively, on the inset of Fig. 3d.

Figure 4(a) depicts the temperature dependence of DC magnetic susceptibility for the studied BiPd single crystal in both Zero Field Cooled (ZFC) and Field Cooled (FC) protocols under applied DC field of 10Oe. A sharp diamagnetic transition is seen at 4K in both ZFC and FC magnetic susceptibility. Isothermal magnetization curve (MH) at temperature 2K is shown in Fig. 4b. The hysteresis loop exhibits typical type-II superconductor behavior. With increasing field the diamagnetism initially keeps on increasing till say 300Oe, starts decreasing for higher fields and become nearly zero at around 500Oe. Seemingly, there is no ferromagnetic signal in the superconducting MH curve at 2K for the studied BiPd single crystal. The lower critical field ($H_{c1}$) has been evaluated from the low field linear part of the MH curve, i.e. Meissner line. The $H_{c1}$ is distinct point on field scale from where the MH deviates by 2% from the perfect Meissner response. Thus seen, $H_{c1}$ for studied BiPd single crystal is depicted in Fig. 4(b). The obtained $H_{c1}(0)$ for BiPd is 287Oe. The MH hysteresis loop at 4K for the BiPd sample is shown in Fig. 4c. Interestingly, the 4K MH loop demonstrates typical ferromagnetic (FM) nature for the studied BiPd. This is interesting because the FM is set in the compound just at the onset of superconductivity. Well, the ferromagnetism in noncentrosymmetric systems could be peripatetic, which is carried by the conduction electrons. It may be due to splitting of the spin-up and spin-down bands. It has been proposed theoretically that the magnetically mediated spin triplet superconductivity and weak ferromagnetism in these systems could be carried by similar electrons [1-3]. Our experimental result being shown in Fig.4c clearly demonstrates that the noncentrosymmetric BiPd superconductor really does have FM component along with superconductivity.

Figure 5 shows the temperature dependence of the electrical resistivity for BiPd single crystal. The data of the electrical resistivity (ρ) were obtained using a standard four-point probe method. With decreasing temperature, ρ display metallic behavior, which becomes zero at below $T_c$ =4K. The normal state resistivity between 5 and 50 K can be described well by a power law $\rho(T) = \rho_o + AT^2$, where $\rho_o$ is the residual resistivity and A the temperature coefficient of the electric resistivity. Upper inset of Fig.5, shows the normalized ρ(T) plot under various applied fields in superconducting temperature range i.e., 2- 5K, Both the $T_c^{onset}$ and $T_c^{(\rho=0)}$ decrease with increasing magnetic field, as expected for a typical type-II superconductor. Lower inset of Fig.5 shows the calculated upper critical field ($H_{c2}$) corresponding to the temperatures, where the resistance drops to 90%, 50% and 10% of the normal state resistivity. The upper critical field of superconductors can be understood as part of two magnetic channels affecting the superconducting state; one the Lorentz-force acting via the charge and the opposite momenta on the paired electrons and another the spin channel i.e., paramagnetic effect. Generally, at higher temperature below $T_c$, the suppression in the orbital channel is more effective and the paramagnetic effect may be visible at much lower temperatures and high fields. Thus, at sufficiently large magnetic fields, the superconductivity is destroyed by both orbital and spin pair breaking. The $H_{c2}(0)$ is estimated by using the conventional one-band Werthamer–Helfand–Hohenberg (WHH) equation, i.e., $H_{c2}(0)=-0.693T_c(dH_{c2}/dT)_{T=Tc}$. The solid lines are the extrapolated to T = 0K, for 90% $\rho_n$ criteria of ρ(T)H. The estimated $H_{c2}(0)$ values for studied BiPd crystal are 7.9kOe, 7.0kOe and 6.8kOe respectively for 90%, 50% and 10% $\rho_n$ criterion. The estimated $H_{c2}(0)$ values for all the BiPd compound is well below the Pauli paramagnetic limit ($\mu_oH_p$=1.84$T_c$), which can be considered as an evidence of a conventional superconducting mechanism.

Fig. 6 shows the temperature dependence of the heat-capacity ($C_p$) of BiPd crystal in temperature range of 2 – 250K. The upper inset of Fig.6 shows the low temperature (below 5K) $C_p/T$ vs $T^2$ plots at various applied fields of 0-2kOe. A clear and distinct jump is seen in $C_p(T)$ at superconducting transition temperature ($T_c$). It is clear that bulk superconductivity sets in BiPd at below 4K. The $C_p$ transition shifts towards the lower temperature side with increasing magnetic field, and the same is not seen down to 2K at applied filed of 2kOe. It is clear that at 2kOe, the $C_p$ of BiPd contains only the electronic part and the phononic part is not seen down to 2K. The electronic part of heat capacity, i.e., the one being measured at 2kOe applied field is fitted to the expression $C_p(T) = \gamma T + \beta T^3 + \delta T^5$. Through the best fitting of specific heat capacity data the coefficients obtained are: $\gamma = 4.41$ mJ mol$^{-1}$ K$^{-2}$, $\beta = 0.91$ mJ mol$^{-1}$ K$^{-4}$ and $\delta = 0.0035$ mJ mol$^{-1}$ K$^{-6}$. Further, with the help of obtained value of $\beta$ the Debye temperature ($\theta_D$) has been calculated by using the relation $\theta_D = (234zR/\beta)^{1/3}$, where R is the Rydberg constant, i.e. 8.314 J mol$^{-1}$ K$^{-1}$ and z is the number of atoms in the BiPd unit cell. The estimated value of $\theta_D$ comes out to be 163K. The lower inset of Fig.6 depicts the change in $\Delta C_p(C_p-C_e)/T$ versus T plot for BiPd. A clear jump $\Delta C_p/T$ is seen at $T_c$ i.e., at below 4K. The magnitude of the jump at $T=T_c$, $\Delta C$, is 6.71mJmol$^{-1}$K$^{-1}$, and the value of the normalized specific-heat jump, $\Delta C/\gamma T_c$, is 1.52. This value is slightly higher than the BCS weak-coupling limit of 1.43. Our value for $\Delta C/\gamma T_c$ of 1.52 for BiPd superconductor is close to that as earlier reported by D. C. Peet et.al [19]. The Kadowaki–Woods ratio i.e., $A/\gamma^2$ is $2.79\times10^{-5}$ $\mu\Omega$ cm mol$^2$ K$^2$ J$^{-2}$, where A is evaluated by fitting of temperature dependent Resistivity in previous section. Interestingly, the value of Kadowaki–Woods ratio for the BiPd sample is suggestive of an intermediate correlated system [20]. Interestingly, in our $C_p(T)$ data (lower inset Fig. 6), two transitions are seen in electronic heat capacity, which may suggest multi gap superconductivity in BiPd compound. This calls for further investigations on nature of superconductivity in BiPd.

Figure 7 shows the temperature dependence of electrical resistivity for the BiPd single crystalline compound at various applied hydrostatic pressures in the temperature range of 300-2K. The normal state resistivity is metallic at ambient pressure as well in hydrostatic pressure up to 2.0GPa. It is clearly seen from Fig.5 that the normal state resistivity increases with increasing pressure from 0-1.0Gpa and the same decreases for the higher pressure of 1.5 and 2.0GPa. Inset of Fig.7 shows the enlarged view of the $\rho(T)$ under pressure in the temperature range 5-2K. The ambient pressure $T_c$ (4K) decreases gradually with increasing pressure, but the decrement is not monotonic. The $T_c$ decrease slowly up to the applied pressure of 1.0GPa, relatively faster for 1.5GPa and becomes nearly saturated for the applied pressure of 2.0GPa. The trend of the decrease in the superconducting transition temperature with applied pressure is shown in Fig. 7b. Liner fitting of these points gives the negative slope for dTc/dP of around 0.62K/Gpa. The decrease in $T_c$ for a low carrier density superconductors can be understood qualitatively in a simple model i.e., $T_c \sim \Theta_D \exp[-1/N(0)V_0]$, where $\Theta_D$ the Debye temperature, $N(0)\sim m^*n^{1/3}$ the density of states and $V_0$ the effective interaction parameter [21]. The increase of normal state resistivity with applied pressure indicates a decrease of the carrier concentration n, which in turn leads to a reduction of $N(0)$ and $T_c$. The negative pressure coefficient of $T_c$ gives strong support for the phonon-mediated BCS-like pairing mechanism for BiPd single crystal compound.

**CONCLUSION**

The investigated crystals of α-BiPd were grown by a self flux melt method. The crystals have a platelet like shape with shiny surfaces. Bulk superconductivity is seen at below 4.0K through DC magnetization and electrical resistivity measurements along with ferromagnetic component. The estimated value of $H_{c2}(0)$ is around 7.0kOe. The value of the normalized specific-heat jump, $\Delta C/\gamma T_c$, is 1.52, which is slightly higher than the BCS weak-coupling limit, of 1.43. The obtained Kadowaki–Woods ratio $(A/\gamma^2)$ of $2.79\times10^{-5}\mu\Omega$cmmol$^2$ K$^2$ J$^{-2}$ is suggestive of an intermediate correlated system. The superconductivity is suppressed with applied pressure, with negative pressure coefficient $dT_c/dP$ of around 0.62K/GPa.

## ACKNOWLEDGEMENT

Authors would like to thank their Director-NPL for his keen interest in the present work This work is supported by DAE-SRC outstanding investigator award scheme to work on search for new superconductors.

**FIGURE CAPTIONS**

**Figures 1:** Room temperature x-ray diffraction patterns of crushed single crystalline BiPd compound; inset is unit cell of BiPd.

**Figures 2:** (a) SEM image of layered BiPd single crystal (b) EDAX graph for the selected area of BiPd single crystal.

**Figures 3:** HRTEM micrographs of BiPd alloy showing (a,b) bright field electron microgrpahs, (c) atomic scale image, and (d) selected area electron diffraction pattern along [110]zone axis of a monoclinic crystal. Insets: (b) atomic scale image of BiPd, (c) atomic scale image of Pd inclusion, and (d) selected area electron diffraction pattern from BiPd and Pd.

**Figures 4:** (a) Temperature dependence of dc magnetic susceptibility in ZFC and FC process for BiPd single crystal, isothermal magnetization versus dc magnetic field at (b) 2K and at (c) 4K for BiPd single crystal.

**Figures 5:** Temperature dependence of electrical resistivity of BiPd single crystal curve fitting in the relation $\rho=\rho_o+ AT^2$, upper inset is $\rho(T)$ under various applied magnetic fields, lower inset is upper critical field ($H_{c2}$) as a function of temperature solid lines is linearly extrapolation of experimental data.

**Figures 6:** Temperature dependence of heat capacity ($C_p$) of BiPd single crystal in temperature range 250-2K. upper inset is $C_p/T$ vs $T^2$ at different fields, lower inset is change of specific heat $C_e/T$ as a function of temperature.

**Figure 7:** (a) Resistivity under different hydrostatic pressure (0-2GPa) as function of temperature for BiPd single crystal, inset shows the magnified view of the same close to superconducting transition. (b) $T_c^{onset}$ vs applied pressure for the BiPd single crystal.

Figure 1

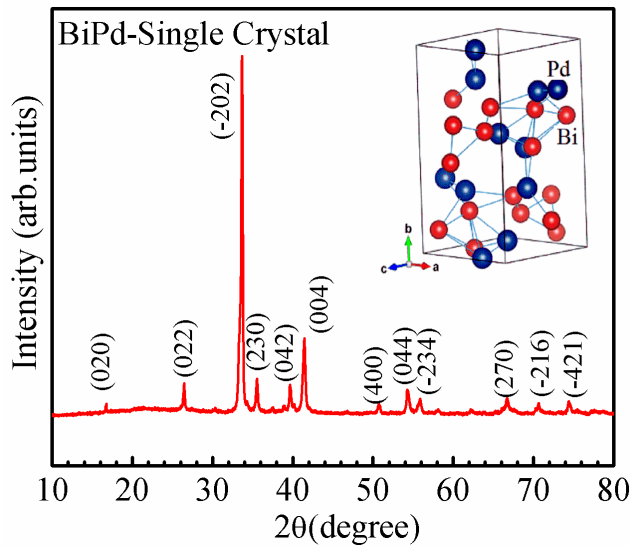

Figure 2

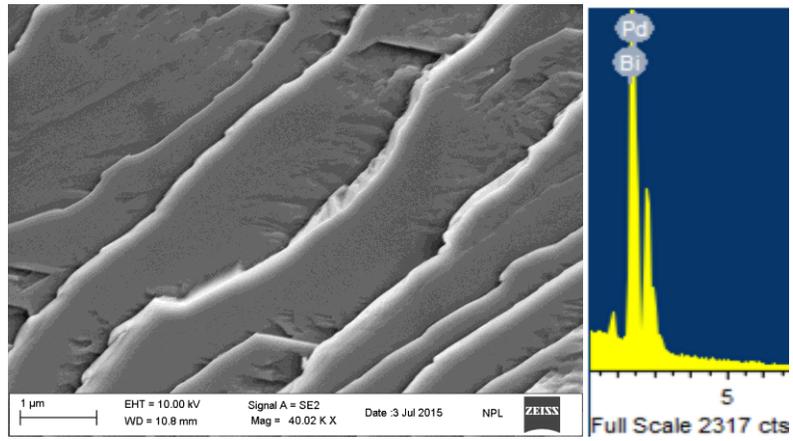

Figure 3

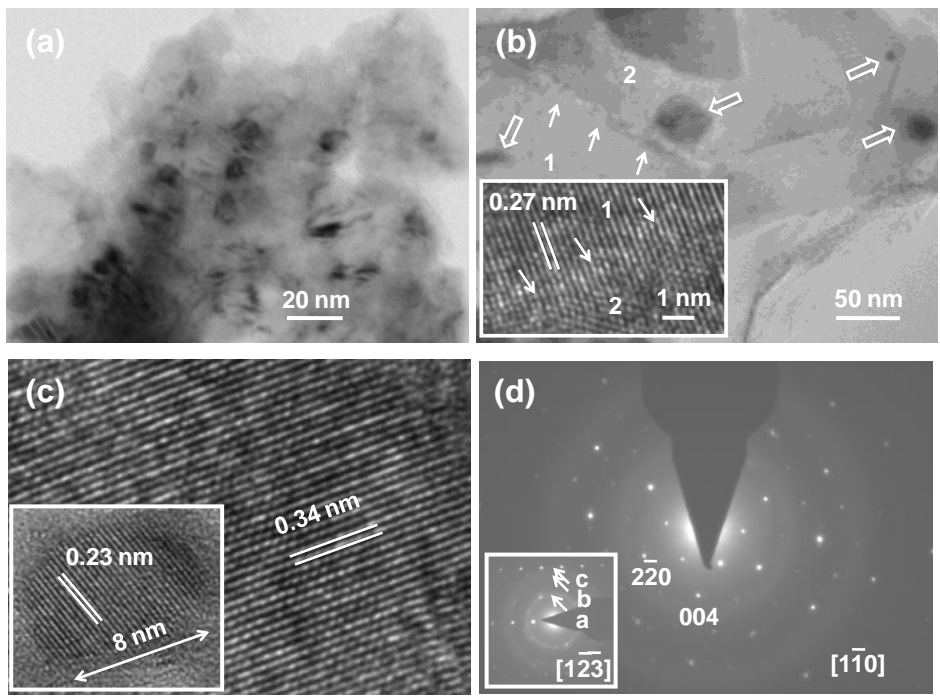

Figure 4

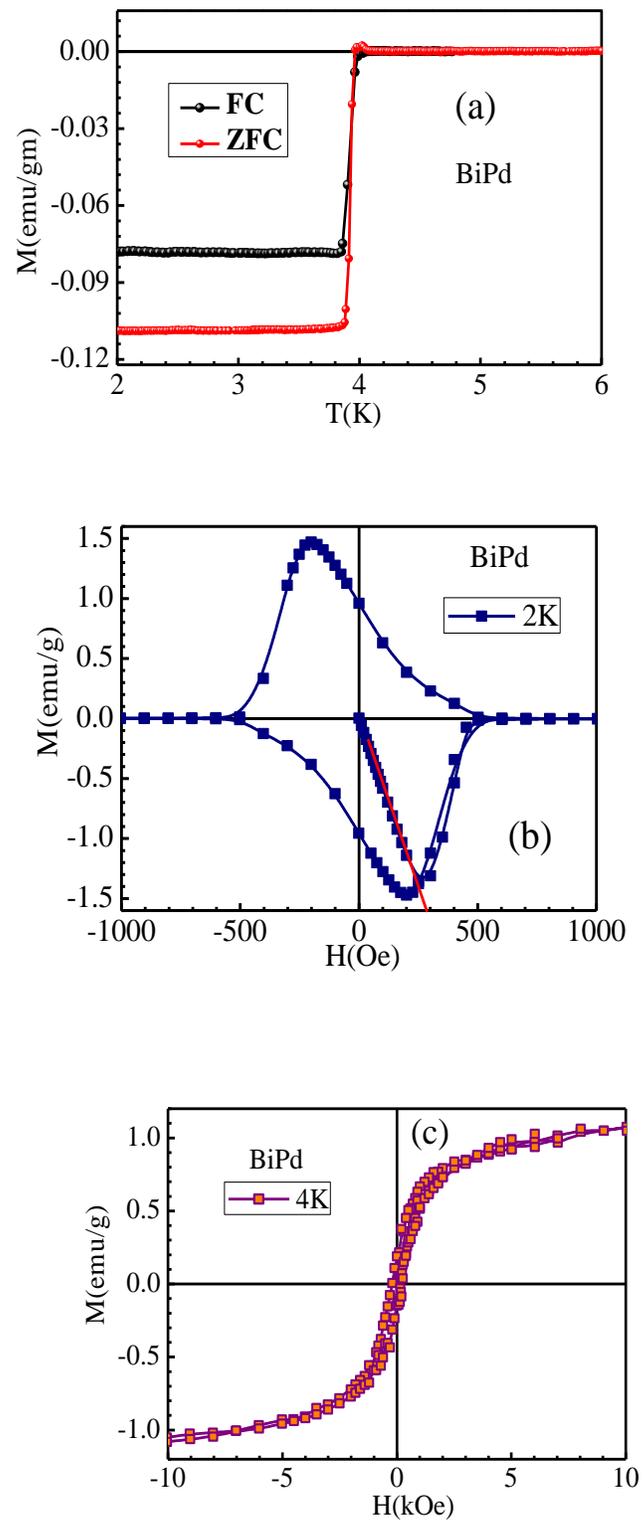

Figure 5

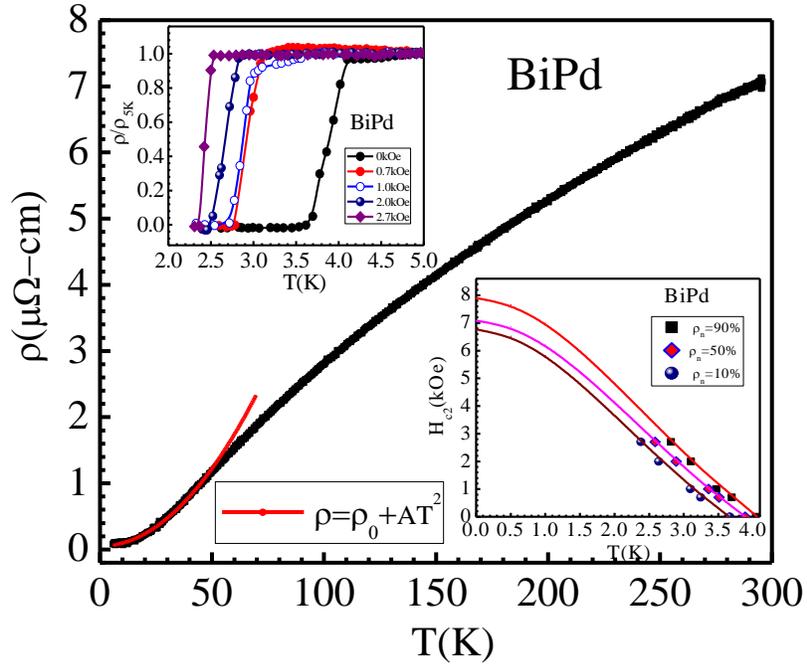

Figure 6

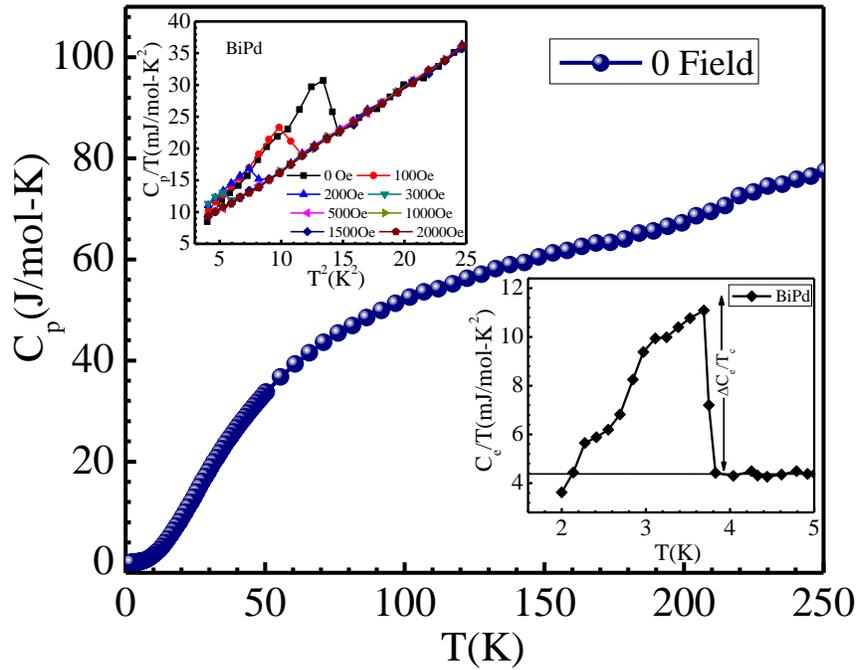

Figure 7

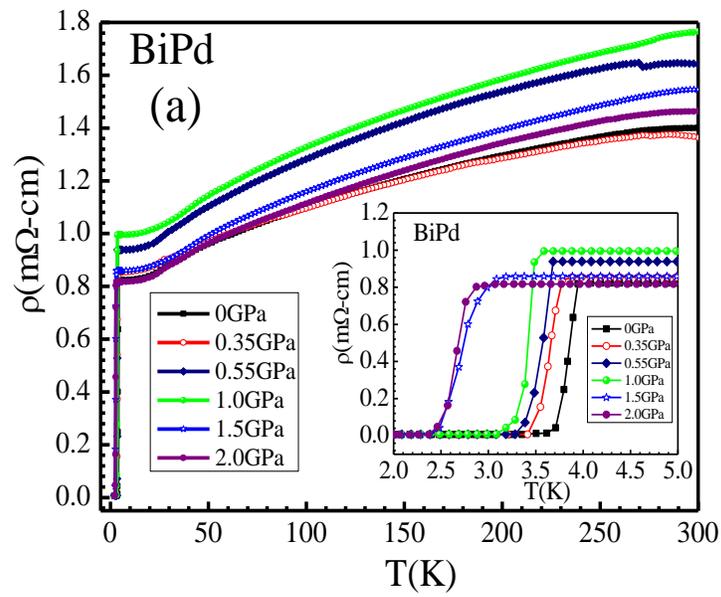

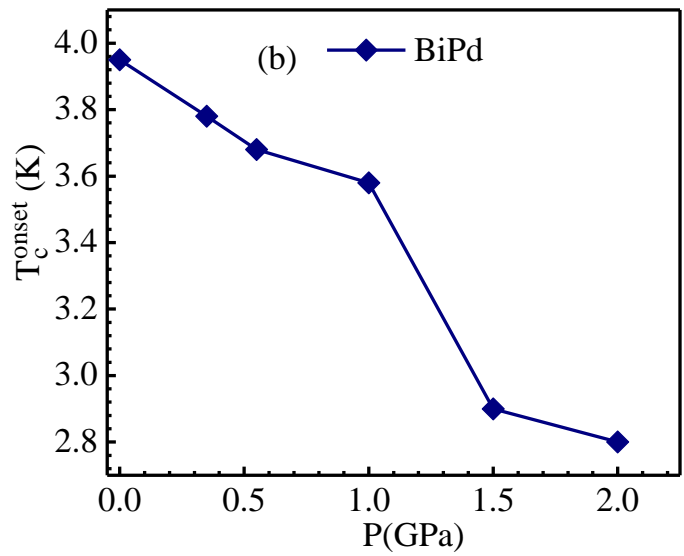